\documentstyle[sprocl,psfig]{article}

\catcode`\@=11
\def\@cite#1#2{{\if@cghi$\!^{#1}$\else$[{#1}]$\fi\if@tempswa\typeout
        {IJCGA warning: optional citation argument
        ignored: `#2'} \fi}}
\catcode`\@=\active

\def\sst{\scriptscriptstyle}
\def\ra{\rightarrow}

\def\nab{\nabla}
\def\del{\partial}

\def\be{\begin{equation}}
\def\ee{\end{equation}}
\def\ba{\begin{array}}
\def\ea{\end{array}}
\def\bea{\begin{eqnarray}}
\def\eea{\end{eqnarray}}
\def\bd{\begin{document}}
\def\ed{\end{document}}
\def\fin{\end{thebibliography}
\end{document}}

\let\la=\label

\let\bm=\bibitem
\def\nn{\nonumber}
\def\qq{\quad\quad}
\let\fr=\frac

\def\ft#1#2{{\textstyle{{\scriptstyle #1}\over {\scriptstyle #2}}}}
\def\fft#1#2{{#1 \over #2}}
\def\sst#1{{\scriptscriptstyle #1}}
\def\oneone{\rlap 1\mkern4mu{\rm l}}
\newcommand{\eq}[1]{(\ref{#1})}
\newcommand{\eqs}[2]{(\ref{#1})-(\ref{#2})}
\newcommand{\w}[1]{\\[0.#1cm]}

\def\Dot#1{\buildrel{_{_{\hskip 0.01in}{\normalsize\bullet}}}\over{#1}}
\def\Hat#1{\widehat{#1}}
\def\bsh{\backslash}
\def\dA{\Dot A}
\def\tA{\tilde{A}}
\def\tZ{\tilde{Z}}
\def\tS{\tilde{S}}
\def\tcF{\tilde{\cal F}}
\def\tcK{\tilde{\cal K}}
\def\dB{\Dot B}
\def\dC{\Dot C}
\def\a{\alpha}
\def\adt{\dot \alpha}
\def\b{\beta}
\def\bdt{\dot \beta}
\def\c{\gamma}
\def\C{\Gamma}
\def\cdt{\dot\gamma}
\def\d{\delta}
\def\D{\Delta}
\def\ddt{\dot\delta}
\def\e{\epsilon}
\def\vare{\varepsilon}
\def\f{\phi}
\def\F{\Phi}
\def\vvf{\f}
\def\h{\eta}
\def\k{\kappa}
\def\l{\lambda}
\def\L{\Lambda}
\def\m{\mu}
\def\n{\nu}
\def\p{\pi}
\def\P{\Pi}
\def\r{\rho}
\def\s{\sigma}
\def\S{\Sigma}
\def\t{\tau}
\def\th{\theta}
\def\Th{\Theta}
\def\vth{\vartheta}
\def\tb{\bar\theta}
\def\X{\Xeta}
\def\x{\xi}
\def\o{\omega}
\def\O{\Omega}

\def\ua{\underline{\alpha}}
\def\ub{\underline{\phantom{\alpha}}\!\!\!\beta}
\def\uc{\underline{\phantom{\alpha}}\!\!\!\gamma}
\def\um{\underline{\mu}}
\def\ud{\underline\delta}
\def\ue{\underline\epsilon}
\def\una{\underline a}
\def\unA{\underline A}
\def\unb{\underline b}
\def\unB{\underline B}
\def\unc{\underline c}
\def\unC{\underline C}
\def\und{\underline d}
\def\unD{\underline D}
\def\une{\underline e}
\def\unE{\underline E}
\def\unf{\underline{\phantom{e}}\!\!\!\! f}
\def\unF{\underline F}
\def\ung{\underline g}
\def\unm{\underline m}
\def\unM{\underline M}
\def\unn{\underline n}
\def\unN{\underline N}
\def\unp{\underline{\phantom{a}}\!\!\! p}
\def\unP{\underline P}
\def\unH{\underline{H}}
\def\unF{\underline{F}}
\def\unT{\underline{T}}
\def\ovA{\overline{A}}
\def\ovB{\overline{B}}
\def\uC{{\underline C}}
\def\ns{\normalsize}
\def\vs{\vspace{-0.25cm}}
\def\se{\;\;=\;\;}
\def\de{\;\;:=\;\;}
\def\cF{{\cal F}}
\def\cH{{\cal H}}
\def\cK{{\cal K}}

\def\atwo{\alpha_{2}}
\def\aone{\alpha_{1}}
\def\afive{\alpha_{5}}
\def\ap{\alpha_p}
\def\azero{\alpha_o}
\def\afour{\alpha_{4}}
\def\appt{\alpha_{p+2}}
\def\apmo{\alpha_{p-1}}

\def\cE{{\cal E}}
\def\tr{{\rm tr}}
\def\bC{{\bar \C}}

\begin{document}

\quad

\vspace{-1in}

\hfill{CTP TAMU-5/99}

\hfill{hep-th/9902171}

\vspace{1in}

\title{ASPECTS OF THE $M5$-BRANE
\footnote{Contribution to the Trieste Conference on Superfivebranes \&
Physics in $5+1$ Dimensions, Trieste, 1-3 April 1998}}

\author{ E. SEZGIN }

\address{Center for Theoretical Physics, Texas A\&M University,
College  Station, \\ TX 77843, USA}

\author{P. SUNDELL}

\address{Department of Engineering Sciences, Physics and Mathematics
Karlstad University, S-651 88 Karlstad, Sweden}

\maketitle \abstracts { The $\k$-symmetry of an open $M2$-brane ending
on an $M5$-brane requires geometrical constraints on the embedding of
the system in target superspace. These constraints lead to the
$M5$-brane equations of motion, which we review both in superspace and
in component (i.e. in Green-Schwarz) formalism. We also describe the
embedding of the chiral $M5$-brane theory in a non-chiral theory where
the equations of motion follow from an action that involves a
non-chiral $2$-form potential, upon the imposition of a non-linear
self-duality condition. In this formulation, we find a simplified form
of the second order field equation for the worldvolume $2$-form
potential, and we derive the nonlinear holomorphicity condition on the
partition function of the chiral $M5$-brane.}

\pagebreak


\section{Introduction}


The $M5$-brane is an important ingredient of $M$-theory. Studies of
coincident $M5$-branes, $M2$-branes stretched between $M5$-branes, and
wrapping of $M5$-branes around various internal space, for example,
have led to discoveries of remarkable non-perturbative phenomena.

While there are many ways to understand the existence of the
$M5$-brane, it was first discovered as a classical solution of $D=11$
supergravity \cite{g}. A particularly interesting way of describing the
$M5$-brane is to view it as the surface on which an open $M2$-brane can
end \cite{str,pkt1}
\footnote{ The open $M2$-brane whose boundary moves freely in target
spacetime was considered briefly in \citelow{bst} but it was realized
that the attendant boundary condition necessarily breaks the $D=11$
Lorentz invariance. Therefore the emphasis was put strictly on the
closed supermembrane in all the early studies of the $M2$-brane.}.
In this picture, among other things, the full $M5$-brane equations of
motion follow from the requirement of $\k$-symmetry of the open
$M2$-brane action.

In this article, we focus on some aspects of the $M5$-brane which deal
with the structure of the $M5$-brane equations of motion and the
embedding of the theory into a non-chiral theory that admits an action
formulation. To be more specific:

\begin{enumerate}

\item[a)] We review the derivation of the basic constraints on the
geometry of the embedding of the $M5$-brane super worldvolume into the
target superspace from the requirement of $\k$-symmetry of an open
$M2$-brane ending on the $M5$-brane \cite{cs}.

\item[b)] We review the key results for the $M5$-brane equations of
motion following from the constraints both in superspace as well as
component (i.e. Green-Schwarz) formalism \cite{hs1,hs2,hsw1}.

\item[c)] We describe the embedding of the chiral $M5$-brane theory in
a non-chiral theory where the equations of motion follow from an action
that involves a non-chiral $2$-form potential, upon the imposition of a
non-linear self-duality condition.

\end{enumerate}

The non-chiral formulation of the $M5$-brane does not contain any
Lagrange multipliers or auxiliary fields. It is, however, equivalent to
a scale invariant formulation \cite{c}, which does contain a Lagrange
multiplier scalar field and a $5$-form potential. It differs, on the
other hand, from the intrinsically chiral formulation of \citelow{s}
which contains an auxiliary scalar field.

In the non-chiral formulation, we also find a simplified form of the
second order field equation for the worldvolume $2$-form potential, and
we derive a nonlinear holomorphicity condition on the partition
function of the chiral $M5$-brane.

The results mentioned in (a) and (b) are covered in Sec. 2 and 3, and
they are based on \citelow{cs} and \citelow{hs1,hs2,hsw1},
respectively. The results mentioned in (c) are covered in Sec. 4, and
they are essentially based on \citelow{c}. Sec. 4 does contain,
however, some new results.


\section{An $M2$-Brane Ending on the $M5$-Brane}


The $M5$-brane equations of motion can be derived from the
considerations of an open $M2$-brane ending on the $M5$-brane
\cite{cs}. Consider an open $M2$-brane ending on the $M5$-brane whose
worldvolume is a $(6|16)$-dimensional supersubmanifold of the
$(11|32)$-dimensional target superspace. We use the notation $(D|D')$,
where $D$ is the real bosonic dimension and $D'$ is the real fermionic
dimension of a supermanifold. Thus, denoting the worldvolume of the
$M2$-brane by $\S$, the $M5$-brane worldvolume by $M_5$ and the target
superspace by $\unM$, we have the chain of embedding:

\be
\del \S \subset M_5 \subset \unM \ .
\la{gsemb}
\ee

In this approach it is important to note the $M2$-brane worldvolume $\S$
is purely bosonic, while the manifold $M_5$ on which it ends, which is of
course the $M5$-brane worldvolume, is a supermanifold. Thus
the worldvolume supersymmetry of the $M5$-brane is manifest, while the
worldvolume supersymmetry of the $M2$-brane is not manifest as it is the
case in any Green-Schwarz type brane action. In this sense this
formulation is a hybrid one.

The worldvolume supersymmetry of the $M2$-brane can be made manifest as
well, by elevating the worldvolume $\S$ into a $(3|16)$-dimensional
supermanifold $M_2$, thus having the superembedding chain: $ \del M_2
\subset M_5 \subset \unM $.

Both approaches yield the same superembedding equations for the
$M5$-brane. These equations, which will be derived below in the hybrid
formulation, are constraints on the embedding that lead to full,
covariant equations for the $M5$-brane that we will spell out in
Section 3.

In this section, we will consider the embedding chain \eq{gsemb}. For
simplicity, we will take $\del\S$ to consist of a single boundary
component. We use the notations and conventions of \citelow{hs1}. In
particular, we denote by $z^{\unM}=(x^{\unm},\th^{\um})$ the local
coordinates on $\unM$, and $A=(a,\a)$ is the target tangent space
index. We use the ununderlined version of these indices to label the
corresponding quantities on the worldsurface. The embedded submanifold
$M$, with local coordinates $y^M$, is given as $z^{\unM}(y)$.

We consider the following action for an open supermembrane ending on a
superfivebrane \cite{cs}

\be
S \se -\int_\S d^3 \xi \left ( \sqrt{-g} + \e^{ijk} C_{ijk}\right)
+ \int_{\del\S} d^2 \s \e^{rs} A_{rs}\ ,\la{action}
\ee

where $\x^i~(i=0,1,2)$ are the coordinates on the membrane worldvolume
$\S$, $\s^r~(r=1,2)$ are the coordinates on the boundary $\del\S$,
$g_{ij}$ is the metric on $\S$ and $g=\det g_{ij}$.

In addition to the usual super 3-form $C_3$ in $(11|32)$ dimensional
target superspace $\unM$, we have introduced a super 2-form $A_2$ on
the $(6|16)$ dimensional superfivebrane worldvolume $M_5$, which is a
{\it supersubmanifold} of $\unM$. The suitable pullbacks of these
superforms, and the induced metric occurring in the action are defined
as:

\bea
C_{ijk} &\de& \del_i z^{\unM}\del_j z^{\unN}\del_k z^{\unP}
C_{\unP\unN\unM}\ ,
\nn\w2
A_{rs} &\de& \del_r y^{M}\del_s y^{N} A_{NM}\ ,
\nn\w2
g_{ij} &\de& \left(\del_i z^{\unM}E_{\unM}{}^{\una}\right)
\left(\del_j z^{\unN}E_{\unN}{}^{\unb}\right)
\eta_{\una\unb}\ ,\la{cbg}
\eea

where $\eta_{\una\unb}$ is the Minkowski metric in eleven dimensions and
$E_{\unM}{}^{\unA}$ is the target space supervielbein.
Defining the basis one-forms $E^{\unA} =
d\xi^i\del_iz^{\unM}E_{\unM}{}^{\unA}$ and
$E^A = d\s^r \del_r y^M E_M{}^A$, where $E_M{}^A$ is the supervielbein on
$M_5$, note the useful relation

\be
E^{\unA}|_{\del\S} \se E^A E_A{}^{\unA}|_{\del\S}\ .\la{useful}
\ee

The embedding matrix $E_A{}^{\unA}$ plays an important role in the
description of the model, and it is defined as

\be
E_A{}^{\unA} \de E_A{}^M\del_{M}z^{\unM}E_{\unM}{}^{\unA},
\ee

The action \eq{action} is invariant under diffeomorphisms of $\S$, with
suitable boundary conditions imposed on the parameter of the
transformation, as well as the tensor gauge transformations $\d C_3= d
\L_2$ and $\d A_2=f_5^*\L_2$ where $\L_2$ is a super 2-form in $\unM$,
and $f_5^\star$ is the pullback associated with the embedding map
$f_5:M_5\hookrightarrow \unM$.

We shall now require the total action to be invariant under the
$\k$-symmetry transformation. On the interior of $\S$, they take the
usual form

\be
\d_{\k} z^{\una} \se 0 \ ,\qq \d_{\k} z^{\ua} \se \k^{\uc}(\xi)
(1+\C_{(2)})_{\uc}{}^{\ua} \ ,
\la{ks1}
\ee

where $ \d_\k z^{\unA}:= \d_\k z^{\unM} E_{\unM}{}^{\unA}$ and

\be
\C_{(2)} \de \frac1{3!\sqrt{-g}} \e^{ijk} \C_{ijk}\ ,
\qq
\C_i \de \del_i z^{\unM}E_{\unM}{}^{\una} \C_{\una}\ .
\ee

We also need to specify the fermionic $\k$-symmetry
transformations of $z^{\unA}$ on the boundary $\del\S$.
Without loss of generality, they take the form

\be
\d_{\k} z^{\una} \se 0\ ,\qq \d_{\k} z^{\ua} \se \k^{\uc}(\s)
P_{\uc}{}^{\ua} \qq \mbox{on $\del\S$} \ ,
\la{ks2}
\ee

where $P_{\uc}{}^{\ua}$ is some projector (see \eq{c5}).

We next derive the consequences of the $\k$-transformations specified
above. To do so, we first observe that an arbitrary transformation of
$y^M$ induces a transformation on $z^{\unM}$ given by

\be
\d z^{\unA} \se \d y^A E_A{}^{\unA}\ \qq \mbox{on $M$},
\la{dzb}
\ee

where $\d y^{A}=\d y^{M} E_{M}{}^{A}$.

It is useful to introduce a normal basis $E_{A'}=E_{A'}{}^{\unA}
E_{\unA}$ of vectors at each point on the worldsurface. The inverse of
the pair $(E_A{}^{\unA},E_{A'}{}^{\unA})$ is denoted by
$(E_{\unA}{}^A,E_{\unA}{}^{A'})$ \cite{hsw1}. It is also useful to
define the projection matrices

\bea
E_{\ua}{}^{\a} E_{\a}{}^{\uc} &\de& \ft12 (1+\C_{(5)})_{\ua}{}^{\uc}\ ,
\nn\w2
E_{\ua}{}^{\a'} E_{\a'}{}^{\uc} &\de& \ft12 (1-\C_{(5)})_{\ua}{}^{\uc}\ ,
\la{p2}
\eea

where $\C_{(5)}$, defined by these equations, satisfies $\C_{(5)}^2=1$.
Its explicit form is not needed at the moment, but it will be spelled
out in the next section (see \eq{af}).

The variation $\d_\k z^{\ua}$ given in \eq{ks2} satisfies $\bar{\k} P
(1-\C_{(5)}) = 0$ on the boundary $\del \S$. This can be seen by
multiplying the $\unA=\ua $ component of \eq{dzb} by $E_{\ua}{}^{\a'}
E_{\a'}{}^{\uc}$ and noting that $E_{\a}{}^{\ua} E_{\ua}{}^{\b'}=0$.
Thus the projector $P$ introduced in \eq{ks2} is given by

\be
P \se \ft12 (1+\C_{(5)})\ .
\la{c5}
\ee

Next we determine $\d_\k y^A$. From \eq{ks2} and \eq{dzb} it follows
that

\be
0\se\d_\k y^a E_a{}^{\una}+  \d_\k y^{\a} E_{\a}{}^{\una}\ ,
\la{dz2}
\ee

on the boundary $\del \S$. The $\una=b$ component of this equation is
$0= \d_\k y^a E_a{}^b+ \d_\k y^{\a} E_{\a}{}^b$. One can check that
$E_\a{}^b$ can be gauged away by using the bosonic diffeomorphisms of
$M$, namely $\d_\eta y^M E_M{}^a=\eta^a$. Hence, one can set
$E_\a{}^b=0$, and since $E_a{}^b$ is invertible, it follows that

\be
\d_\k y^a\se 0\ , \la{dza0}
\ee

on $\del\S$, and hence on $M$. Next, using \eq{dza0} in \eq{dzb}, we
find $ \d_\k y^\a E_\a{}^{\ua}= \d_\k z^{\ua}$ on the boundary
${\del\S}$ which implies $ \d_{\k} y^{\a}= \d_\k z^{\ua}
E_{\ua}{}^{\a}$ on the boundary ${\del\S}$. This means that the
variation $\d_\k y^\a$ is an arbitrary odd-diffeomorphism, effecting
the 16 fermionic coordinates of $M$, and that when restricted to
$\del\S$, it agrees with the $\kappa$-symmetry transformation on
${\unM}$, which also has 16 independent fermionic parameters.

We now turn to the derivation of the constraint on the $M5$-brane
embedding mentioned earlier. Using this in $\una=b'$ component of
\eq{dz2}, and observing that $\d_\k y^\a$ is an arbitrary odd
diffeomorphism of $M$, it follows that $E_\a{}^{b'}=0$. Recalling that
$E_\a{}^b=0$ as well, we get

\be
E_{\a}{}^{\una}\se 0\ .
\label{basic}
\ee

This is the superembedding condition that plays a crucial role in the
description of superbrane dynamics \cite{hs1,hs2,hsw1}.

Now we are ready to seek the conditions for the $\k$-symmetry of the
action \eq{action}. Using \eq{dzb} and \eq{dza0} in the variation of
the action, we find that the vanishing of the terms on $\S$ imposes
constraints on the torsion super 2-form $T^{\unA}$ and the super 4-form
$H_4=dC_3$, such that they imply the equations of motion of the eleven
dimensional supergravity \cite{bst}. The non-vanishing parts of the
target space torsion are \cite{cf,bh}

\bea
T_{\ua\ub}{}^{\unc} &\se& -i(\C^{\unc})_{\ua\ub}\ ,
\nn\w2
T_{\una\ub}{}^{\uc} &\se&-
{1\over36}(\C^{\unb\unc\und})_{\ub}{}^{\uc}H_{\una\unb\unc\und}
-{1\over288}(\C_{\una\unb\unc\und\une})_{\ub}{}^{\uc}
H^{\unb\unc\und\une}\ ,
\la{tt}
\eea

and $T_{\una\unb}{}^{\uc}$. The only other non-vanishing components of
$H_4$ are

\be
H_{\una\unb\uc\ud}\se -i(\C_{\una\unb})_{\uc\ud}\ .
\la{h4}
\ee

The remaining variations are on the boundary, and yield the final result

\be
\d_\k S\se\int_{\del \S} \e^{rs} \left(\del_r y^M E_M{}^A\right)
\left(\del_s y^N E_N{}^B\right)\d_\k y^\c  \cF_{\c B A}\ ,
\la{var}
\ee

where we have introduced the following super $3$-form in $M_5$:

\be
\cF_3\de dA_2-f_5^{*} C_3\ .
\la{cf3}
\ee

Since $\d_\k y^\a$ are arbitrary, the vanishing of \eq{var} implies the
constraint

\be
\cF_{\c B A}\se 0\ .
\la{h}
\ee

Thus the only non-vanishing component of $H$ is $H_{abc}$. The
constraints \eq{basic} and \eq{h} encode elegantly all the information
on the superfivebrane dynamics, as has been shown in
\citelow{hs1,hs2,hsw1}.

Finally we consider the boundary conditions that arise from the
variation of the action \eq{action}. The requirement of the action be
stationary when the supermembrane field equations of \citelow{bst} hold
can readily be shown to impose the following mixed Dirichlet and Neumann
boundary conditions

\be
\d z^{a'}|_{\del\S} \se 0\ ,\qq \left( \sqrt{-g}
n^iE_{ic}+n_i\e^{ijk} E_j{}^a E_k{}^b H_{abc}\right)|_{\del\S}\se 0\ ,
\la{dbc}
\ee

where $n^i$ is a unit vector normal to the boundary $\del\S$, and $a'$
labels the directions transverse to the fivebrane worldvolume. The
reparametrization invariance of \eq{dbc} imposes the boundary condition
$n^i \del_i v^r|_{\del\S} = 0$ and the reparametrization invariance of
the leads to the further boundary condition $ n_i v^i |_{\del\S} = 0$.


\section{The Covariant $M5$-Brane Equations of Motion}


Here, we give the nonlinear field equations of the superfivebrane
equations, up to second order fermionic terms, that follow from the
superembedding condition $E_\a{}^{\una}=0$, which are proposed to arise
equally well from the ${\cF}$-constraint $\cF_{\a BC}=0$. The details
of the procedures can be found in \citelow{hsw1}. A key point is the
emergence of a super $3$-form $h$ in world superspace. This form arises
in the following component of the embedding matrix

\be
E_{\a}{}^{\ua} \se u_{\a}{}^{\ua}+h_{\a}{}^{\b'} u_{\b'}{}^{\ua}\ ,
\la{s}
\ee

where, upon the splitting of the indices to exhibit the $USp(4)$
R-symmetry group indices $i=1,...,4$, we have

\be
h_{\a}{}^{\b'}\ra h_{\a i \b}{}^{j}\se
{1\over 6} \d_i{}^j(\c^{abc})_{\a\b} h_{abc}\ ,
\ee

where $h_{abc}$ is a self-dual field defined on $M$. The pair
$(u_{\a}{}^{\ua},u_{\a'}{}^{\ua})$ make up an element of the group
$Spin(1,10)$.

The superembedding formalism was shown to give the following complete
$M5$-brane equations of motion:

\bea
&&E_a{}^{\ua}E_{\ua}{}^{\b'}(\C^a)_{\b'}{}^{\a}\se 0\ ,
\nn\w2
&&\eta^{ab}\nabla_a E_n{}^{\una} E_{\una}{}^{b'}\se -\ft18 (\C^{b'a})_{\c'}{}^{\b}
Z_{a\b}{}^{\c'}\ ,
\nn\w2
&&\hat\nabla^c h_{abc}\se -\ft1{32}(\C^c\C_{ab})_{\c'}{}^{\b}Z_{c\b}{}^{\c'}\ ,
\la{se1}
\eea

where

\be
Z_{a\b}{}^{\c'}\se E_{\b}{}^{\ub}\left(E_{a}{}^{\una} T_{\una\ub}{}^{\uc}
-E_{a}{}^{\ud}E_{\ub}{}^{\c} (\nabla_{\c}E_{\ud}{}^{\d'})E_{\d'}{}^{\uc}
\right)E_{\uc}{}^{\c'} \ .
\la{z}
\ee

Recall that the inverse of the pair $(E_A{}^{\unA},E_{A'}{}^{\unA})$ is
denoted by $(E_{\unA}{}^A,E_{\unA}{}^{A'})$ and that $A=(a,\a)$ label
the tangential directions while $A'=(a',\a')$ label the normal
directions to the $M5$-brane worldvolume.

The target space torsion components $T_{\una\ub}{}^{\uc}$ are given in
\eq{tt} and the second term involves only quantities that are bilinear
in worldvolume fermions. The covariant derivative $\hat\nabla$ has an
additional, composite $SO(5,1)$ connection of the form $(\nabla
u)u^{-1}$ as explained in more detail in \citelow{hsw1}.

The $M5$-brane equations of motion \eq{se1} live in superspace
\cite{hs1,hs2}. The component (i.e. Green-Schwarz) form of these
equations have also been worked out \cite{hsw1}. Up to fermionic
bilinears, the final result is:

\bea
&& {\cE}_a(1-\C)\c^b m_b{}^a \se 0\ ,
\nn\w2
&& G^{mn}\nab_m
{\cF}_{npq}\se Q^{-1}\left[4Y-2(mY+Ym)+mYm\right]_{pq}\ ,
\la{e1}\w2 &&
G^{mn}\nab_m{\cE}_n{}^{\unc} \se {Q\over \sqrt{-g}} \e^{m_1\cdots m_6
}\left(\ft1{6!}H^{\una}{}_{m_1\cdots m_6} + \ft1{(3!)^2}
H^{\una}{}_{m_1m_2m_3}\,{\cF}_{m_4m_5m_6}\, \right)P_{\una}{}^{\unc}\ .
\nn
\eea

Several definitions are in order. To begin with,

\bea
&& m_a{}^b \de \d_a{}^b-2 k_a{}^b\ ,\qq
k_a{}^b \de h_{acd} h^{bcd}\ ,\qq Q \de (1-\ft23 \tr\,k^2)\ ,
\nn\w2
&& Y_{ab} \de \left[4\star H-2(m\star H+\star H m)+m\star H m\right]_{ab}\ ,
\nn\w2
&& P_{\una}{}^{\unc} \de
\d_{\una}{}^{\unc}-{\cE}_{\una}{}^m{\cE}_m{}^{\unc}\ ,
\quad\quad \star H^{ab} \de \ft1{4!\sqrt{-g}}\e^{abcdef}H_{cdef}\ ,
\la{defs}
\eea

The fields ${\cF}_{abc}$, $H_{\una_1\cdots\una_4}$ and its Hodge dual
$H_{\una_1\cdots\una_7}$ are the purely bosonic components of the
superforms

\be
{\cF}_3 \se dA_2-{\unC}_3\ , \qq H_4 \se dC_3 \ , \qq H_7 \se dC_6
+ \ft12\,C_3\wedge H_4\ .
\la{h7}
\ee

The remaining nonvanishing component of $H_7$ is

\be
H_{\ua\ub\una\unb\unc\und\une}\se
-i(\C_{\una\unb\unc\und\une})_{\ua\ub}\ .
\la{hh7}
\ee

The target space indices on $H_4$ and $H_7$ have been converted to
worldvolume indices with factors of ${\cE}_m{}^{\una}$ which are the
supersymmetric line elements defined as

\bea
{\cE}_m{}^{\una}(x) &\de& \del_m z^{\unM} E_{\unM}{}^{\una}\qquad
{\rm at}\ \th=0\ ,
\nn\w2
{\cE}_m{}^{\ua}(x) &\de& \del_m z^{\unM}
E_{\unM}{}^{\ua}\qquad {\rm at}\ \th=0\ .
\eea

The metric

\be
g_{mn}(x) \de {\cE}_m{}^{\una}{\cE}_n{}^{\unb}\h_{\una\unb} \se
e_m{}^a e_n{}^b \eta_{ab}
\ee

is the standard $GS$ induced metric with determinant $g$, and $G^{mn}$
is another metric defined as

\be
G^{mn} \de (m^2)^{ab}e_a{}^m e_b{}^n\ .
 \la{gmn}
 \ee

Let us note that the connection in the covariant derivative $\nabla_m$
occurring in \eq{e1} is the Levi-Civita connection for the induced
metric $g_{mn}$ up to fermionic bilinears.

A key relation between $h_{abc}$ and ${\cF}_{abc}$ follows from the
dimension-$0$ components of the Bianchi identity $d{\cF}_3=-{\unH}_4$,
and is given by
\footnote{We have rescaled the $\cF_3$ of \citelow{hsw1} by a factor of
$4$.}

\be
h_{abc}\se \ft14\,m_a{}^d {\cF}_{bcd}\ .
\la{hf}
\ee

The matrix $\C$ is the $\th =0$ component of the matrix $\C_{(5)}$
introduced above in \eq{p2} and it is given by

\be
\C \se -\bC+\ft13 h^{mnp}\C_{mnp} \se
- \left[{\rm exp}~ (-\ft13\C^{mnp}h_{mnp})\right] \bC\ ,
\la{af}
\ee

where

\bea
&& \bC \de {1\over 6!\sqrt {-g}} \e^{m_1\cdots m_6}
\C_{m_1\cdots m_6}\ ,
\la{c0}\w2
&& \C_m\de {\cE}_m{}^{\una}\C_{\una} \ ,
\qq \C^b\de\C^m e_m{}^b\ , \qq e_m{}^b \de E_m{}^a m_a{}^b\ .
\nn
\eea

The $\k$-symmetry transformation rules are

\bea
\d_{\k} z^{\una} &\se& 0 \ ,
\nn\w2
\d_{\k} z^{\ua} &\se&  \k^{\uc}\ft12 (1+\C)_{\uc}{}^{\ua} \ ,
\nn\w2
\d_\k h_{abc} &\se& -\ft{i}{16} m_{d[a}\,{\cE}^d(1-\C)\C_{bc]} \k\ ,
\la{kt}
\eea

where $\C$ is given by \eq{af}. The $\k$-symmetry transformations are
the fermionic diffeomorphisms of the $M5$-brane worldvolume with
parameter $\k^\a=\k^{\ua}E_{\ua}{}^{\a}$. Thus, using \eq{h}, it
follows immediately that

\be
\d_{\k} \cF_3\se \{d,i_\k\}\cF_3\se -i_\k \unH_4\ ,
\la{kcf}
\ee

which can also be verified by direct computation by combining \eq{hf}
and \eq{kt}.

The equations of motion \eq{e1} have been shown \cite{s} to be
equivalent to those which follow from an action with auxiliary scalar
field \cite{s2}.

We conclude this section by elucidating the consequences of the central
equation \eq{hf}. To this end, we first note the useful identities

\bea
h_{abe}h^{cde}&\se&\d^{[c}_{[a}k_{b]}^{\ \;d]}\ ,
\nn\w2
k_{ac}k_b{}^c&\se&\ft16 \eta_{ab}\tr~k^2\ ,
\nn\w2 k_a{}^d
h_{bcd}&\se&k_{[a}{}^d h_{bc]d}\ ,
\la{hk1}
\eea

which are consequences of the linear self-duality of $h_{abc}$. Taking
the Hodge dual of \eq{hf} one finds $\star\cF_{abc} = -\cF_{abc} + 2
Q^{-1} m_a{}^d \cF_{bcd}$. Using the identity $m^2=2m-Q$, we readily
find the nonlinear self-duality equation

\be
\star \cF_{mnp} \se Q^{-1}G_m{}^q \cF_{npq}\ .
\la{sd}
\ee

This equation can be expressed solely in terms of $\cF_3$. To do this, we
first insert \eq{hf} into \eq{hk1}, which yields the identities

\bea
\cF_{abe}\cF^{cde} &\se& 2\delta_{[a}^{[c} X_{b]}^{\ \;d]} + \ft12
K^{-2} X_{[a}{}^{c} X_{b]}{}^{d} + 2(K^2-1) \delta_{[a}^{c}
\delta_{b]}^{d}\ ,
\nn\w2
X_{ac}X_b{}^c &\se& 4K^2(K^2-1)\eta_{ab}\ ,
\nn\w2
X_a{}^d \cF_{bcd} &\se& X_{[a}{}^d \cF_{bc]d}\ .
\la{x}
\eea

where we have defined

\bea
K &\de& \sqrt{1+\ft1{24}\cF^{abc}\cF_{abc}}\ ,
\la{k}\w2
X_{ab} & \de & \ft12 K \star \cF_a{}^{cd}\cF_{bcd}\ .
\la{xx}
\eea

Next we derive the identities

\bea
Q(K+1) &\se& 2\ ,
\nn\w2
X_{ab} &\se&  \ft12 \cF_{acd}\cF_{b}{}^{cd} -
\ft1{12} \eta_{ab} \cF_{cde}\cF^{cde} \se 4 K(1+K) k_{ab}\ .
\eea

We can now express \eq{sd} entirely in terms of $\cF_3$
by deriving the identity

\be
Q^{-1}G_{mn} \se K \eta_{mn} - \ft12 K^{-1} X_{mn} \ .
\ee

Another way of writing \eq{sd} is

\be \cF^-_{abc}\se\ft12 (1+K)^{-2} \cF^+_{ade}\cF^{+def}\cF^+_{fbc}\ ,
\ee

where $K$ is a root of the quartic equation

\be
(K+1)^3(K-1)\se \ft1{24} \cF^{+ abc}\cF^{+}_{ade}\cF^{+def}\cF^+_{fbc}\ .
\ee


\section{The $M5$-brane Action for an Unconstrained $2$-Form Potential}


While the superembedding constraints yield the covariant equations of
motion it is desirable to have an action from which these equations of
motion can be derived. There exists a universal action formula which
emerges naturally in the superembedding approach \cite{hrs2}. However,
as we shall see in Section 5, this action formalism is not directly
applicable to branes with self-dual field strengths in the worldvolume
such as the $M5$-brane.

A manifestly target space supercovariant and $\k$-invariant action,
which contains an auxiliary scalar and from which the self-duality
condition can be derived as an equation of motion has been constructed
\cite{s}. However, as has been argued by Witten, any attempt to even
define a proper partition function for the $M5$-brane using this action
requires a choice of the auxiliary scalar whose topological class in
general breaks some of the symmetries of $M$-theory. The root of this
problem lies in the fact that the theory, in effect, describes a chiral
$2$-form \cite{w}.

One resolution of this problem involves the embedding of the chiral
theory into non-chiral one \cite{w}. At the classical level, this amounts to
finding an action involving an unconstrained $2$-form potential $A_2$
such that its field equation is equivalent to the Bianchi identity
$d\cF_3=-\unH_4$ and the action is $\k$-invariant upon the imposition
of the self-duality condition.

At the quantum level, the decoupling of the unwanted chirality
components amounts to imposing a constraint on the partition function
$Z[C_3]$, which, at the linearized level, reads \cite{w}

\be
D_{abc} Z[\unC_3] \se -i\sqrt{-g} \left<\cF^+_{abc}\right>\ ,
\la{dz}
\ee

where the functional derivative $D_{abc}$ is defined as

\be
 D_{abc}\de {\d\over \d C^{abc}} + \ft{i}2 \sqrt{-g} \star C_{abc}\ .
 \la{d}
\ee

Thus $D^-_{abc}Z=0$, and only $\cF^+_{abc}$ couples to $C_{abc}$.

Motivated by these considerations, a non-chiral extension of the
$M5$-brane was constructed in \citelow{c}, where it was found that the
connection between $\k$-symmetry and self-duality is sufficient to
determine the action and the non-linear self-duality condition \eq{sd}.

The key to the construction of the action is the fact that the
non-linear self-duality condition \eq{sd} can be written in the
following form

\be
\star \cF_3\se {\del\cK\over\del \cF_3}\ .
\la{delk}
\ee

where $\cK$ can be computed with the help of the identities presented
in the previous section. The result is

\be
\cK \se  2\;\sqrt{1+\ft1{12}\cF^2+\ft1{288} (\cF^2)^2 -\ft1{96}
                  \cF_{abc}\cF^{bcd}\cF_{def}\cF^{efa}}\ ,
\la{ck}
\ee

modulo terms whose $\cF_3$-derivatives vanish when \eq{sd} holds. One
also finds the useful relation \cite{c}:

\be
\cK \se 2 K \ \qq \mbox{for }\quad \star \cF_3 \se { \del \cK \over
\del \cF_3 } \ .
\la{ckk}
\ee

In view of \eq{delk}, a suitable action is given by

\be
S\se \int ( \ft12 \star\cK - Z_6 )\ ,
\la{s1}
\ee

where the Wess-Zumino term

\be
Z_6 \se {\unC}_6- \ft12 {\unC}_3\wedge {\cF}_3\ .
\la{z6}
\ee

The rational behind this action is as follows: treating $\cF_3$ as
subject only to the Bianchi identity $d\cF_3=-\unH_4$, and varying
\eq{s1} with respect to the {\it unconstrained} two-form potential
$A_2$ one finds that\footnote{Notice that the $C_6$ term in the action
does not affect the $A_2$ field equations. Nonetheless it is needed for
$\kappa$-symmetry of the action.}

\be
d\left(\star{\del\cK\over\del\cF_3}\right)\se -\unH_4\ .
\la{ste}
\ee

Combining this second order equation for $A_2$ with the Bianchi
$d\cF_3=-\unH_4$, we find that the only possible self-duality condition
that can be imposed is precisely \eq{delk}. Moreover it was shown in
\citelow{c} that the action \eq{s1} has the $\k$-symmetry characterized
by

\be
\d_\k Z^{\unM} \se \k^{\unM}\ , \qq \d_\k A_2 \se i_\k \unC_3\ ,
\la{ks3}
\ee

provided that the self-duality condition \eq{sd} is satisfied and the
$\k$-parameter is projected as

\be
\k^{\unM}\se \k^{\ub}\ft12(1+\C')_{\ub}{}^{\ua}\,E_{\ua}{}^{\unM}\ ,
\la{ks4}
\ee

where

\be
\C' \se K^{-1}\left(-\bC+\ft1{12}\star\cF^{mnp}\C_{mnp}\right)\ .
\la{kc}
\ee

The equivalence between the $\k$-symmetry transformations
\eqs{ks3}{ks4} and those which arise from the superembedding formalism
as given in \eqs{kt}{kcf}, follows from the identities

\be
(1+\C')(1-\C)\se 0\se (1+\C)(1-\C')\ ,
\la{gg}
\ee

which can be shown with the help of \eq{hf} and \eq{x}.

We emphasize that the self-duality condition \eq{sd} does not follow
directly as an equation of motion from the action \eq{s1}. Instead, as
we saw above, it is recovered as the only self-dual truncation of the
theory that is consistent in the sense that it interchanges the Bianchi
identity $d\cF_3=-\unH_4$ with the tensor field equation \eq{ste}.
Actually, the form of $\cK$ and the self-duality condition \eq{sd} can
also be understood \cite{c} by starting from an action of the form
$S=\int(\ft12 \star \cK-Z_6)$ and demanding invariance under the
$\k$-symmetry transformations of the form \eqs{ks3}{ks4}.

We can now derive the non-linear version of the constraint \eq{dz} by
starting from the action \eq{s1} and the formal definition

\be
Z[\unC_3] \se \int DA_2 ~e^{iS}\ .
\ee

Using the functional derivative \eq{d} we get

\bea
D_{abc}e^{iS}&\se & \ft{i}2\sqrt{-g}
\left({\del \cK\over \del C^{abc}} -  \star(dA_2)_{abc}  +
\star C_{abc}\right)e^{iS}
\nn\w2
&\se & -\ft{i}2\sqrt{-g} \left( {\del \cK\over \del \cF^{abc}} +
\star \cF_{abc}\right)e^{iS}\ .
\eea

which means that

\be
D_{abc}Z[\unC_3]\se  -\ft{i}2\sqrt{-g} \left< {\del \cK\over \del \cF^{abc}} +
                 \star \cF_{abc}\right>\ .
\ee

Since the right side is a projection onto the nonlinearly self-dual
part of $\cF_3$, this is a proper generalization of the constraint
\eq{dz} to the nonlinear case. The full consequences of this constraint
remain to be investigated.


\subsection*{\bf The $M5$-Brane Equations of Motion}


The $\k$-symmetry transformations \eqs{ks3}{ks4} map the non-linear
self-duality condition \eq{delk} into the $z^{\unM}$ equations of
motion, which therefore must agree with the corresponding results
\eq{e1} obtained from the superembedding. It is nonetheless instructive
to demonstrate the equivalence of the field equations by direct
computation. The equations of motion following from the action \eq{s1}
followed by the use of the self-duality equation \eq{sd} are:

\bea
&& \cE_m J^m\se 0 \ ,
\la{d2}\w2
&& G^{pq}\nabla_{p}\cF_{qmn} \se-Q\left(\star H_{mn} +
\ft12 \cF_{mnp}\cF^{pqr}\star H_{qr}\right)\ ,
\nn\w2
&& G^{mn}\nab_m{\cE}_n{}^{\unc} \se{Q\over \sqrt{-g}}
\e^{m_1\cdots m_6 }\left(\ft1{6!} H^{\una}{}_{m_1\cdots m_6}
+\ft1{(3!)^2} H^{\una}{}_{m_1m_2m_3}\,{\cF}_{m_4m_5m_6}\,
\right)P_{\una}{}^{\unc}\ ,
\nn
\eea

where the symmetric bispinor $J^m$ is given by

\be
J^m\se \C^m\bC + \ft12 \star \cF^{mnp}\C_{np} - Q^{-1}G^{mn}\C_n\ .
\la{jm}
\ee

The $z^{\unA}$ field equation arises as an admixture of the variation
with respect to $z^{\unA}$ and the tensor field equation obtained by
varying the action with respect to $\d z^{\unA}:=V^{\unA}$ and $\d
A_2=i_V \unC_3$, such that $\d \cF_3 = -i_V \unH_4$.

In obtaining the field equations \eq{d2}, it is useful to realize that
$Q^{-1}G_{mn}$ actually is the energy-momentum tensor associated with
the composite metric $g_{mn}$:

\be
{\d S_{kin}\over \d g_{mn}}\se \ft12 \sqrt{-g}~ Q^{-1}G^{mn}\ ,
\la{tmn}
\ee

where $S_{kin}=\int \ft12\star \cK$. The invariance of $S_{kin}$ under
the bosonic worldvolume diffeomorphism $\d g_{mn}=2\nabla_{(m}\x_{n)}$
and $\d \cF_3 = di_\x A_2 - i_\x \unH_4$ then implies

\be
\nabla_n (Q^{-1}G^{mn})\se -\ft16 H^{mnpq}\star \cF_{npq}\ ,
\la{div}
\ee

provided that \eq{ste} holds.

Next, we compare the equations of motion \eq{d2} with \eq{e1} obtained
from superembedding. The scalar field equations are already in the same
form. To show equivalence of the Dirac and tensor equations to those
obtained in superembedding formalism requires some work. Let us begin
with the Dirac equation. The $\k$-symmetry of this equation implies
that

\be
(1+\C')J^m\se 0 \se J^m(1-\C'^T)\ .
\la{ks}
\ee

Eq. \eq{gg} then implies

\be
(1+\C)J^m\se 0 \se J^m(1-\C^T)\ .
\ee

From \eq{af} it follows that $\C=-X^{-1}\bC X$ and $\C^T=-X\bC X^{-1}$
where we have introduced $X:=\exp(\ft16 h^{abc}\C_{abc})$, which
implies that $(1-\bC) X J^m = 0 = J^m X (1+\bC)$. On the other hand,
from $m^{ab}\C_{b}\bC=-\bC m^{ab}\C_{b}$ and $(1-\C)\bC=1-\C$ it
follows that the Dirac equation in \eq{e1} is annihilated from the right
by $1+\bC$. Hence, upon multiplication from right by $X$ the Dirac
equation \eq{d2} will be proportional to \eq{e1}. Indeed one can verify
that $ (1-\C)m^{ab}\C_b= -Q J^a X$, which shows the equivalence between
the two Dirac equations.

As for the tensor field equation, we have verified that it turns into
the one obtained in the superembedding formalism (see \eq{e1}) by using
the identity $m^2=2m-Q$. It is worth noting that the field equation for
$A_2$ given in \eq{d2} is considerably simpler than the form it takes
in \eq{e1}.


\subsection*{Dualization of the Non-Chiral Action }


As a check of the formalism, let us briefly discuss the dualization of
the non-chiral theory. To this end we introduce a dual $2$-form
potential $\tA_2$ as a Lagrange multiplier for the Bianchi identity
$d\cF_3=-\unH_4$ and integrate out $A_2$ via its field strength
$\cF_3$, which yields the dual partition function

\be
\tZ[\unC_3]\se \int D\cF_3\; D\tA_2 ~e^{iS-\ft{i}2 \int \tA_2\wedge
d(\cF_3+\unC_3)} \de \int D\tA_2 ~ e^{i\tS}\ .
\ee

The dual action $\tS$ can easily be computed in the saddle point
approximation which yields

\be
\tS\se \int (\ft12 \star \tcK - \tZ_6)\ ,
\la{ts} \ee

where we have used the notations $\,\star \tcK:=\star \cK-\tcF_3\wedge
\cF_3$ and $\tZ_6:= \unC_6-\ft12 \unC_3\wedge \tcF_3$ where $\cF_3$ is
supposed to be expressed in terms of the dual field strength $\tcF_3$
via

\be
\tcF_3\de d\tA_2-\unC_3\se \star{\del \cK\over \del \cF_3}\ .
\la{tcf}
\ee

Varying \eq{ts} with respect to $\tA_2$ and using \eq{tcf}, one finds
the dual second order tensor field equation $d\cF_3=-\unH_4$, where, as
mentioned above, $\cF_3$ is expressed in terms of $\tcF_3$ via
\eq{tcf}. Thus, in the saddle point approximation, the chiral
truncation of the dual theory is given by $\tcF_3=\cF_3$, which from
\eq{tcf} is seen to be equivalent to the condition \eq{delk} for chiral
truncation of the original non-chiral theory. In other words, the
non-chiral theory and its dual have equivalent chiral truncations.


\subsection*{\bf A Scale Invariant Formulation of the Non-Chiral
Action}


The action discussed at length in the previous section is equivalent to
the scale invariant form of the action constructed in \citelow{c}. In
the latter formulation, a worldvolume Lagrange multiplier scalar field
$\l$, and a worldvolume $5$-form potential $A_5$, with field strength
$\cF_6=dA_5+Z_6$, are introduced. This construction is parallel to the
scale invariant formulations of super $p$-brane actions that has been
known for sometime \cite{pkt2,eb1}.

The $M5$-brane action constructed in \citelow{c} is given by

\be
S'\se \ft12 \int \star \l \left(\ft14 \cK^2 - (\star \cF_6)^2
\right)\ .
\la{1}
\ee

To verify that the equations of motion following from this action are
equivalent to those which follow from \eq{s1}, we begin by varying
\eq{1} with respect to $\l$ and the $5$-form potential $A_5$. This
yields two first order field equations, namely $\star \cF_6= \pm \ft12
\cK$ (where the sign reflects the duality of the $2$-form potential),
which determines $A_5$ up to a gauge, and $d( \l \star \cF_6 ) = 0 $,
which we solve by taking $\l= 2T_5 \, \cK^{-1}$, where the integration
constant $T_5$ is the $M5$-brane tension. The remaining equations of
motion from $S'$ follow by varying $z^{\unM}$ and the $2$-form
potential $A_2$. Denoting a general variation of this kind by $\d$, we
find that

\be
\d S' \se\ft12 \int \l \left[\star\left( \Big(\ft14\cK^2 +(\star
\cF_6)^2\Big){1\over \sqrt{-g}}\d\sqrt{-g}\,+\,\ft12 \cK\,\d\cK\right)
-2(\star \cF_6) \d Z_6\right]\ .
\la{13}
\ee

Using the relations $\star \cF_6= \ft12 \cK$ and $\l =2T_5 \, \cK^{-1}$
in this formula, one then immediately finds $ \d S' = T_5\d S$.


\subsection*{\bf Relation to an $M5$-Brane Action in
                 Superembedding Approach}


It has been shown how Green-Schwarz type actions can be systematically
constructed for most branes starting from the superembedding approach
\cite{hrs2}. The construction yields a general action formula. The only
branes for which this formula runs into an obstacle are those which
contain worldvolume chiral $p$-form potentials, such as the $M5$-brane.
It is, nonetheless, interesting to see the result one obtains by a
naive application of this action formula to this case. In doing so, we
will find an action which is closely related to the one discussed
above.

The application of the action formula to the $M5$-brane proceeds as
follows. Defining

\be
W_7 \;\;:=\;\; dZ_6 \se {\unH}_7 + \ft12 {\unH}_4\wedge {\cF}_3
\la{w7}
\ee

and using the fact that de Rham cohomology of the supermanifold $M$
coincides with that of its body $M$ one can always write

\be
W_7\se d K_6
\la{dk}
\ee

for some {\it globally} defined $6$-form $K_6$ on $M$. Furthermore,
since none of the target space fields or the worldsurface fields has
negative dimension, it follows that the only non-vanishing component of
$K_6$ is the purely bosonic one. In components this means

\be
K_{\a A_1 \cdots A_5}\se 0\ .
\la{ka}
\ee

The application of the general action formula of \citelow{hrs2} to the
present case gives the functional

\be
S'' \;\;:=\;\; \int_{M_0}i^*(K_6-Z_6)\ ,
\ee

where $i:M_0 \hookrightarrow M$ is the embedding of the body $M_0$ of
$M$ into $M$. $S''$ is by construction only defined for self-dual
$h_{abc}$ or, equivalently, for the $3$-form field strength $\cF_{abc}$
obeying the non-linear self-duality condition \eq{delk}. $S''$ is
manifestly invariant under reparametrizations of $M_0$ and the
$\k$-transformations \eq{kt} and \eq{kcf} as these transformations are
generated by $i^*v$, where $v$ is a supervectorfield on $M$, and
$\d_{i^*v}i^*L_6=i^*\{d,i_v\}L_6=di^*i_vL_6$ where $L_6=K_6-Z_6$.

To find $K_6$ we insert the constraint \eq{ka} into \eq{dk}. After some
algebra one finds that the dimension $0$ component yields

\be
\star K_6 \se K\ ,
\ee

where $K$ is given by \eq{k}. Therefore, in view of \eq{ckk}, $S''$ is
simply the restriction of the action \eq{s1} to the constraint surface
defined by the non-linear self-duality condition \eq{delk}.

\bigskip\bigskip

\noindent{\large \bf Acknowledments}

\medskip

It is a pleasure to thank our collaborators P.S. Howe, P.C. West,
B.E.W. Nilsson and M. Cederwall. One of the authors (E.S.) would like
to thank the Abdus Salam International Center for Theoretical Physics
for making the {\it Trieste Conference on Superfivebranes \& Physics in
$5+1$ Dimensions} possible and for hospitality. This research has been
supported in part by NSF Grant PHY-9722090.

\vfill\eject


\section*{References}


\end{document}